# HARMONI@ELT: instrument design, performance, and science capabilities

B. Neichel[a*], J. Dunlop[b], S. Chittick[c], D. Le Mignant[a], M. Swinbank[d], A. de Lorenzo-Cáceres[e], W. Taylor[c], M. Tecza[f], B. Johnson[c], A. Costille[a], N. Bouché[g], A. Delsanti[a], T. Morris[d], J. Piqueras Lopez[h], B. Garcia[e], M. Mateo[i], A. Carlotti[j], Dan Dickens[c], Edgard Castillo Dominguez[f] , and the HARMONI Consortium

[a]LAM, Aix Marseille Univ., CNRS, CNES, France; [b]Univ. of Edinburgh, UK; [c]UK Astronomy Technology Centre, STFC, UK ; [d]Univ. of Durham, UK; [e]Instituto de Astrofísica de Canarias, Tenerife, Spain ; [f]University of Oxford, UK ; [g]Centre de Recherche en Astrophysique de Lyon, [h]Centro de Astrobiología (CAB), CSIC-INTA, Madrid, Spain ; [i]University of Michigan, USA ; [j]Institut de Planétologie et d'Astrophysique de Grenoble ;

[1] Benoit.neichel@lam.fr

## ABSTRACT

HARMONI is the integral field spectrograph for the Extremely Large Telescope (ELT). Designed as the primary workhorse near-infrared spectrograph of the ELT, it will deliver spatially resolved spectroscopy at diffraction-limited angular resolution over the wavelength range 0.75–2.4 µm, addressing science questions ranging from the characterization of exoplanet atmospheres to the internal kinematics of galaxies at the epoch of reionization. This paper presents the current status of the HARMONI project following its recent rescope, describing the new instrument architecture, its key subsystems, and the three adaptive optics modes (Single Conjugate AO (SCAO), Multi-Conjugate AO (MCAO) provided by MORFEO, and High Contrast AO (HCAO)). We present the top-level technical specifications and the expected on-sky performance in terms of Strehl ratio, PSF FWHM, sky coverage, sensitivity, and astrometric accuracy. We describe the new project organization, with a new leadership and Project Management Office (PMO), and provide an update on the project schedule and the path toward instrument delivery.

## 1. INTRODUCTION

The Extremely Large Telescope (ELT), currently under construction by ESO on Cerro Armazones in the Atacama Desert of Chile, will be the world's largest optical-near-infrared telescope with a primary mirror diameter of 39 meters. When it enters operations, by 2030, the ELT will offer an angular resolution approximately 6 times finer than the James Webb Space Telescope (JWST) in the H band, and a collecting area that is several hundred times larger than any existing ground-based facility. Unlocking this transformational capability requires a new generation of instruments that can fully exploit the telescope aperture while operating at or close to the diffraction limit.

HARMONI is the designated integral field spectrograph for the ELT. It is designed as the primary workhorse spectrograph of the telescope, providing spatially resolved spectroscopy over the wavelength range 0.75-2.4 µm at two spectral resolutions and two spatial scales, under diffraction-limited correction delivered by three distinct adaptive optics modes.

In the past 2 years, HARMONI has undergone a rigorous rescoping process, co-led by the consortium and ESO, that has delivered a simplified, compliant, and technically robust instrument design.[1,2] The new design focuses the instrument on its core scientific strengths, preserves and in several cases enhances the most critical capabilities, and provides a credible and achievable path to delivery.

This paper is structured as follows. Section 2 describes the science drivers and key use cases that have shaped the rescoped instrument design. Section 3 presents the instrument architecture and key subsystems. Section 4 describes the adaptive optics modes and expected AO performance. Section 5 presents the expected scientific performance, including sensitivity and astrometric stability. Section 6 describes the project organisation and schedule.

## 2. SCIENCE DRIVERS AND KEY USE CASES

HARMONI's scientific programme is organised around six themes, each addressed by a dedicated Science Working Group (SWG). The underlying design philosophy is to ensure the two most transformational capabilities of an IFU at the ELT are protected: diffraction-limited imaging with high sky coverage, and intermediate-resolution spectroscopy over the full near-infrared wavelength range. These two capabilities, in combination, position HARMONI to address science questions that are simply inaccessible to any other facility currently in operation or under construction.

### 2.1 Solar System

Within the Solar System, HARMONI will deliver spatially resolved near-infrared spectroscopy of planetary surfaces, atmospheres, rings, and satellite systems at unprecedented angular resolution and sensitivity. Key targets include the ice giants Uranus and Neptune, for which the spatial resolution of the ELT permits the mapping of albedo features, atmospheric bands, and ring structures at scales inaccessible to current facilities; the surfaces of large Trans-Neptunian Objects, for which the detection of surface ices and organic materials at high spatial resolution provides constraints on the early solar system; and the volcanic activity of Io, monitored through thermal emission maps at sub-arcsecond scales.

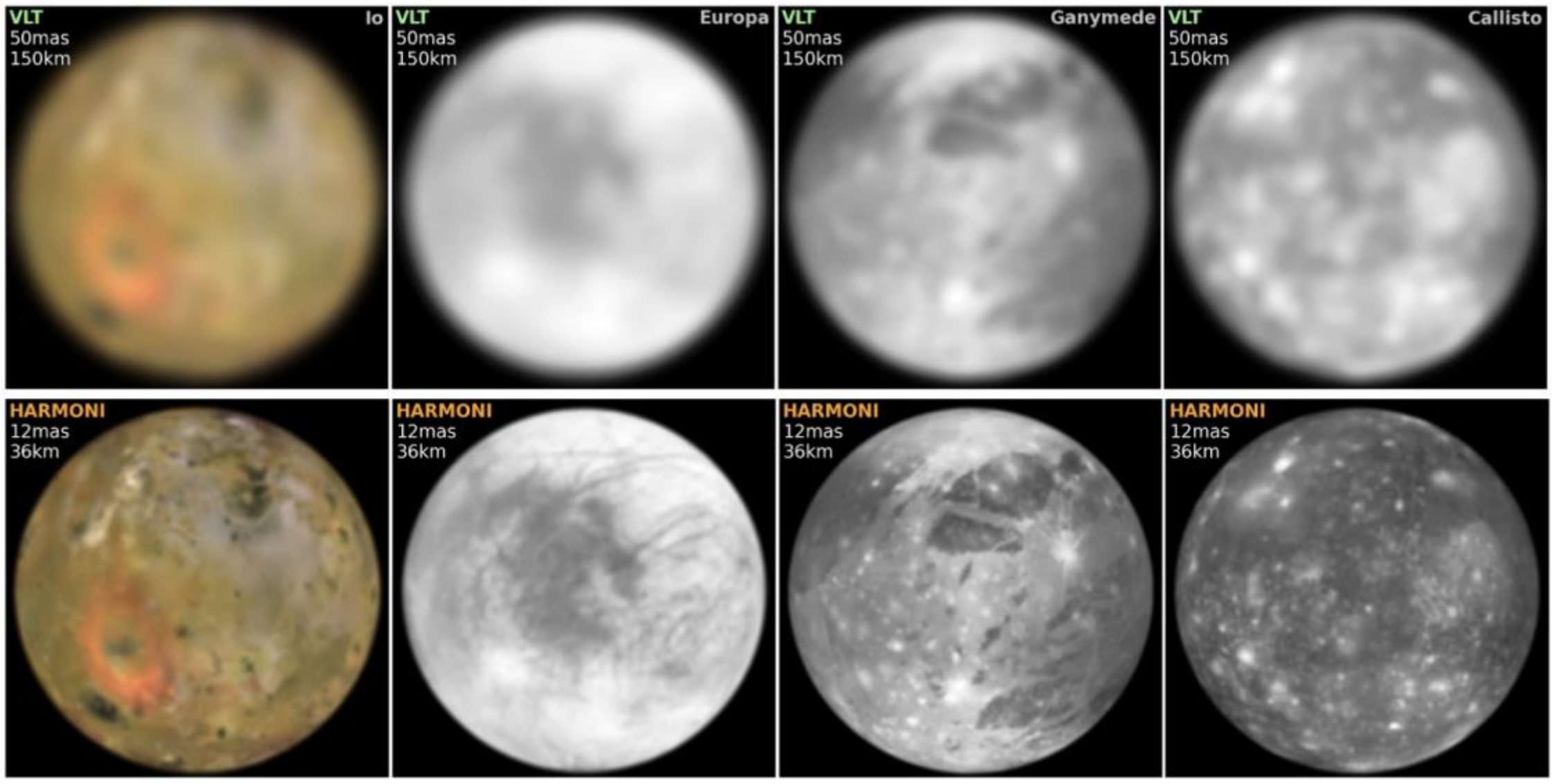

*Figure 1. Simulated HARMONI images of the Galilean satellites (Io, Europa, Ganymede, Callisto) compared to current VLT imaging (50 mas), illustrating the gain in spatial resolution (12 mas) enabled by the ELT for surface characterisation of Solar System bodies. Credit : Olivier Groussin for the Solar System Science group.*

### 2.2 Resolved stellar populations

Beyond the Local Group, the study of stellar populations in external galaxies has historically been limited by angular resolution: stellar crowding prevents the separation of individual stars in all but the most favourable environments, forcing observers to work with integrated light and its inherent degeneracies. HARMONI breaks this barrier. At the distance of the Virgo cluster (~17 Mpc), the ELT diffraction limit in the H band (approximately 10 mas, corresponding to ~0.8 pc) enables the resolution of individual stars on the upper asymptotic giant branch, providing age and metallicity diagnostics that are not accessible through integrated spectroscopy. With a sensitivity gain of several orders of magnitude compared to 8 m-class telescopes operating in the near-infrared, HARMONI will deliver individual stellar spectra across a representative sample of galaxy morphologies, environments, and evolutionary histories, providing the first truly direct view of the chemical and kinematic enrichment history of nearby galaxies.

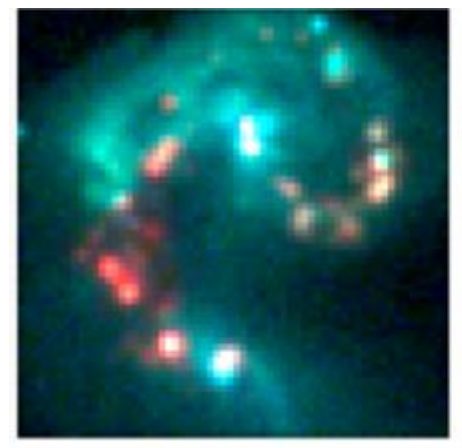
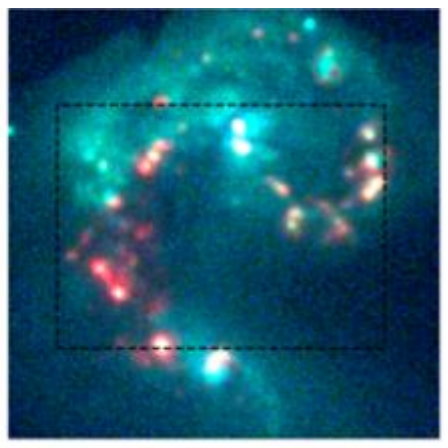

*Figure 2. Simulated HARMONI observations of the Antennae Galaxies (NGC 4038/4039) at the 25 mas (left) and 6 mas (right) spatial scales, demonstrating HARMONI's ability to resolve individual star-forming regions and study them in detail. Credit: Floriane Leclercq.*

### 2.3 Nearby galaxies and supermassive black holes

HARMONI's combination of angular resolution and sensitivity makes it an unrivalled tool for the study of the nuclei of nearby galaxies. The sphere of influence of a supermassive black hole of mass M• scales as $r^H \sim GM•/\sigma^2$, where σ is the stellar velocity dispersion. For the most massive black holes in the Virgo and Fornax clusters (M• ~ $10^9$ M☉, σ ~ 300 km/s), $r^H$ subtends approximately 100 mas at the cluster distance, comfortably resolved by the ELT. More crucially, HARMONI will access the sphere of influence of lower-mass black holes at closer distances and in galaxies with lower velocity dispersions, dramatically expanding the census of dynamically confirmed black hole masses and enabling stringent tests of the M•–σ relation and co-evolution scenarios. [3]

### 2.4 Galaxy formation at high redshift

Among the most compelling capabilities of HARMONI is the prospect of obtaining spatially resolved near-infrared spectra of galaxies at the peak of cosmic star formation (z ~ 2–3) and into the epoch of reionisation (z > 6). At these redshifts, the rest-frame optical emission lines that are the primary diagnostics of star formation rate, metallicity, excitation state, and gas kinematics are redshifted into the J, H, and K bands, precisely the spectral range covered by HARMONI. With angular resolutions of about 10 mas at these wavelengths, HARMONI resolves structures on scales of 50–100 pc at z ~ 2, comparable to the sizes of individual star-forming clumps, HII regions, and compact bulges. This spatial resolution, combined with a sensitivity that puts HARMONI 50–100 times faster than JWST/NIRSpec for point sources at the same spectral resolution, makes HARMONI uniquely capable of resolving the internal physics of high-redshift galaxies that JWST can only detect at the ensemble level. [4]

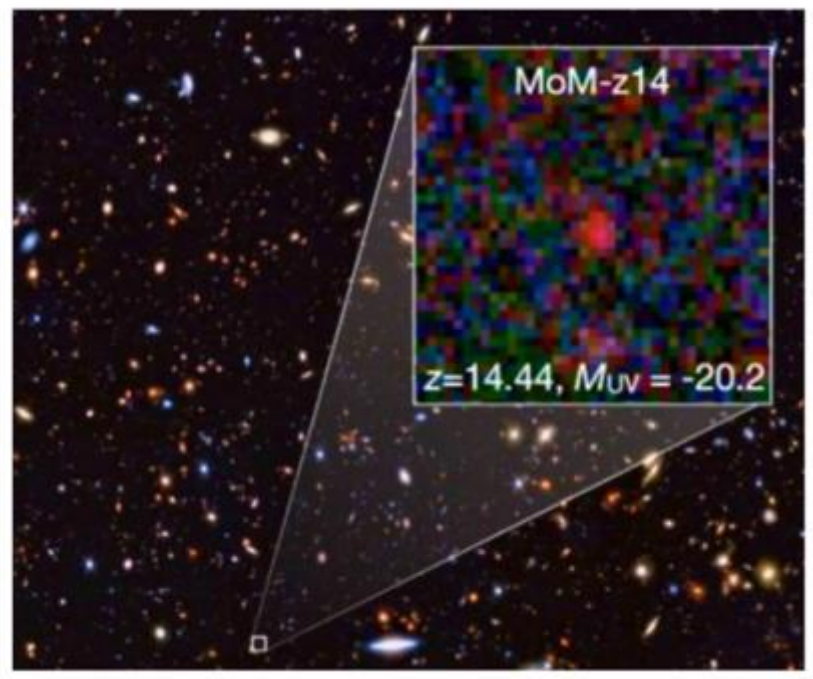

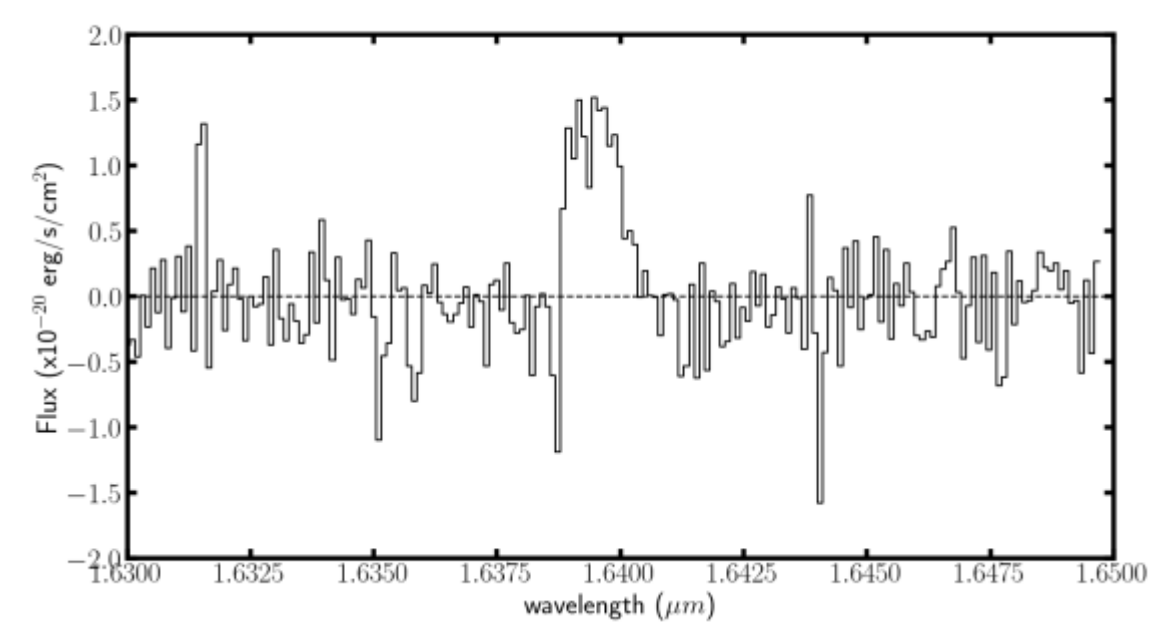

*Figure 3. Simulated HARMONI detection of a Population III star cluster signature in a z~14 galaxy (based on MoM-z14), showing the emission-line detectability achievable in 5-hour for a flux of $5x10^{-19}$ erg/s/cm$^2$ at the epoch of reionisation [5]*

### 2.5 Exoplanet detection and characterisation

In the High Contrast AO mode (Section 4.3), HARMONI will directly image and spectrally characterise giant exoplanets in the H and K bands at angular separations of 50–500 mas. The 3.9 mas spaxel scale provided in HCAO mode, combined with coronagraphic apodizers and focal plane masks, enables the suppression of stellar diffraction light to a contrast of $10^{-6}$ at 100 mas separation in a single 60-second exposure for bright guide stars. This performance enables the detection of Jupiter-mass planets at projected separations of 1–2 AU around stars at distances of 20 pc, the regime corresponding to the water snow line for solar-type stars, where giant planet formation is expected to be most efficient. Beyond detection, integral field spectroscopy of the planet in H+K bands enables molecular mapping through cross-correlation techniques, providing direct constraints on the C/O ratio, atmospheric temperature structure, and abundance of key molecules such as CO, $H_2O$, and $CH_4$.

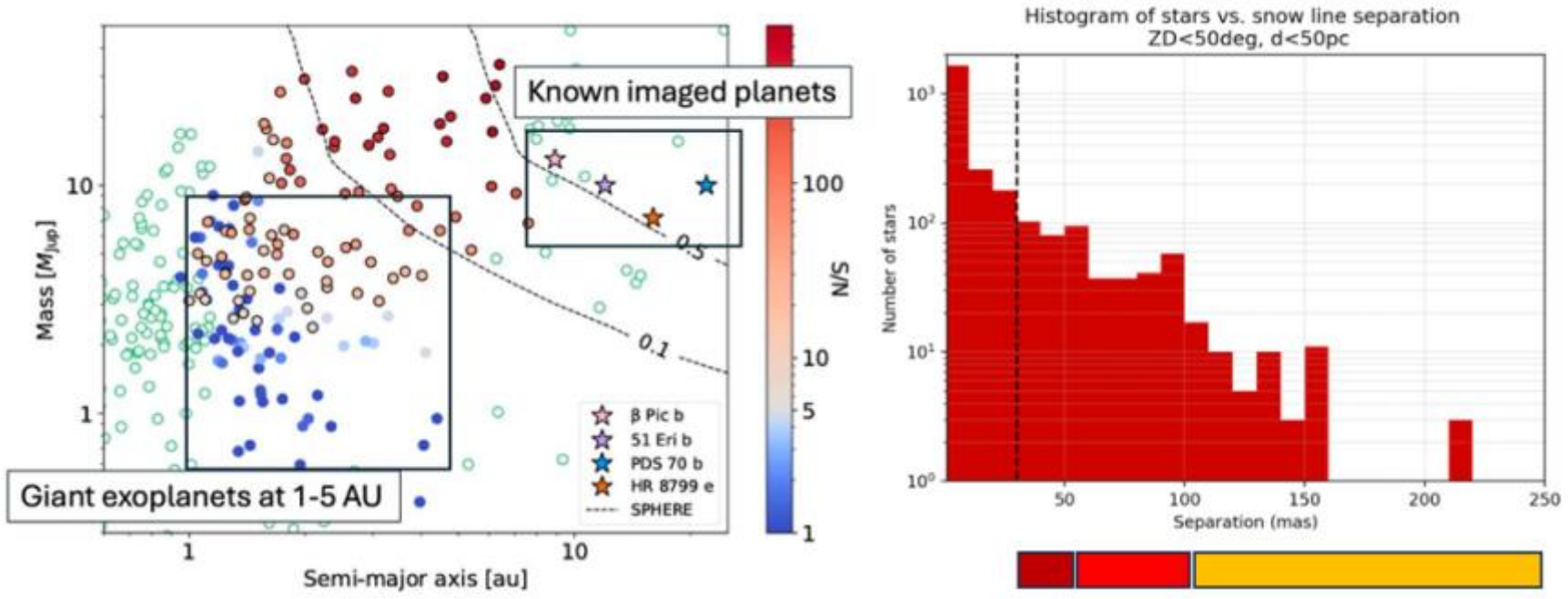

*Figure 4. Left: known directly imaged exoplanets and giant planets at 1–5 AU in the mass–separation plane (adapted from Houllé et al. 2021); right: histogram of snow-line separations for young stars within 50 pc (MOCA database, Gagné et al. 2026).*

### 2.6 Transient universe

The time-domain capabilities of HARMONI, combined with the ability to obtain spatially resolved spectra of transient sources and their immediate environments, make it a powerful tool for the follow-up of explosive transients. For kilonova events associated with neutron star mergers detected by gravitational wave observatories, HARMONI will characterise the r-process nucleosynthesis products through near-infrared spectroscopy at the very location of the event, providing the physical conditions and yields that are inaccessible from unresolved photometric light curves. The blue wavelength extension of the LR1 grating to 0.75 µm (described in Section 3.3) was motivated in part by this science case, enabling the full coverage of the P-Cygni features in kilonova spectra that carry the chemical fingerprint of r-process elements.

## 3. INSTRUMENT DESIGN AND ARCHITECTURE

The rescoped HARMONI is a near-infrared integral field spectrograph operating over the wavelength range 0.75–2.4 µm. The instrument is fed by three adaptive optics modes and provides two spatial scales and two spectral resolution families, delivering six distinct spectral configurations. The architecture has been designed around two guiding principles: scientific completeness across the two core capabilities, and technical robustness that can be reliably delivered within the available resources and schedule.

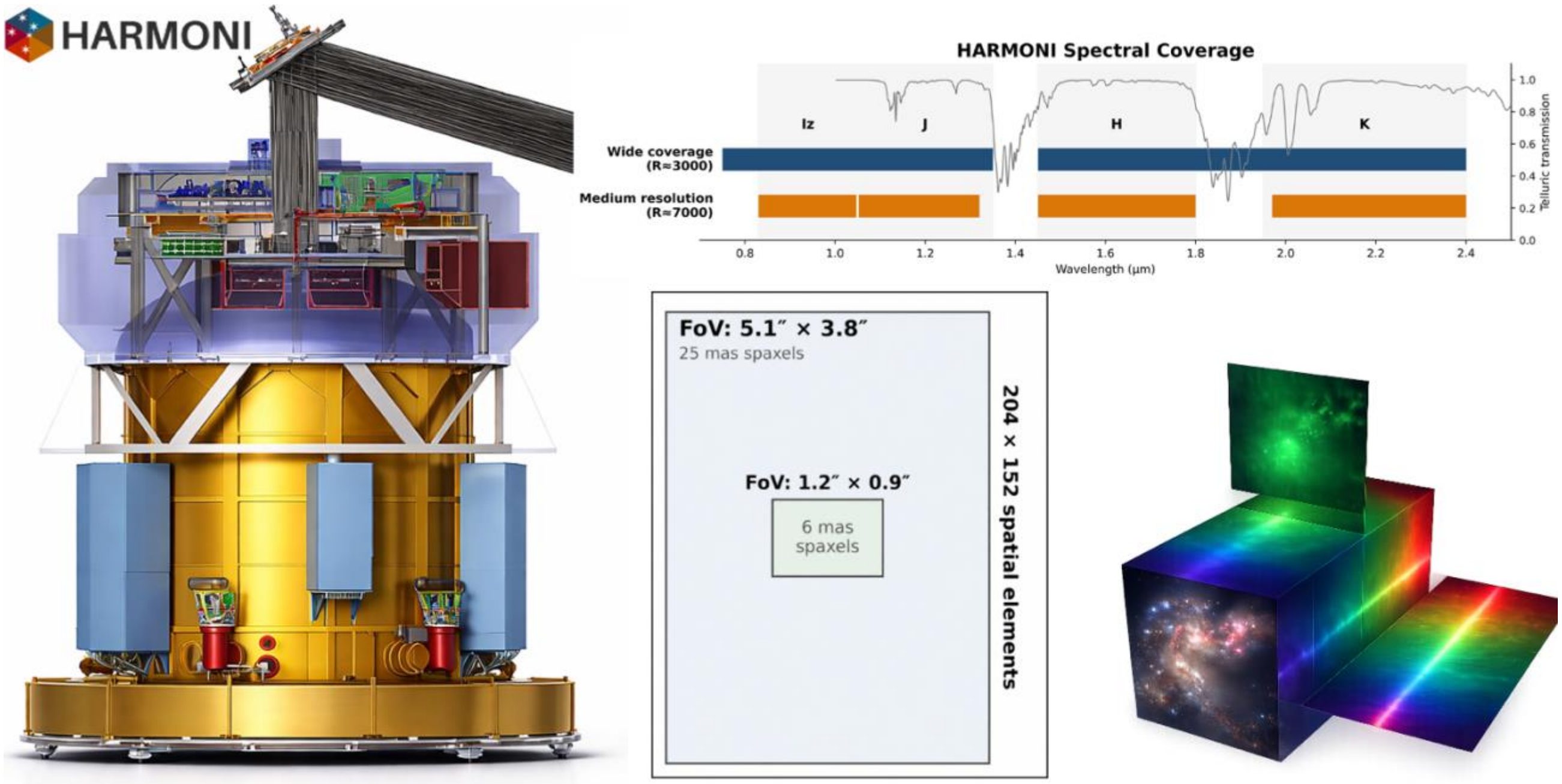

*Figure 5. HARMONI spectral coverage in the wide (R≈3000) and medium-resolution (R≈7000) settings across the Iz, J, H, and K bands (top), and the two spatial scales and corresponding fields of view delivered by the IFS (bottom).*

The overall instrument architecture consists of three main functional blocks: (1) the Calibration and Relay System (CARS), which relay and reformat the focal plane from the ELT Nasmyth platform; (2) the IFS cold assembly, housed inside a cryostat operating at liquid nitrogen temperatures, comprising the Image Slicer Module, the Splitting and Relay Module, and four spectrograph channels; and (3) the Natural Guide Star Sensing System (NGSS). The design reduces the number of motorized cryogenic mechanisms from 26 to 14 (including 4 that are set at integration and never moved during operations), and the number of achievable instrument configurations from 181 to 26, dramatically simplifying the risk landscape relative to earlier concepts.

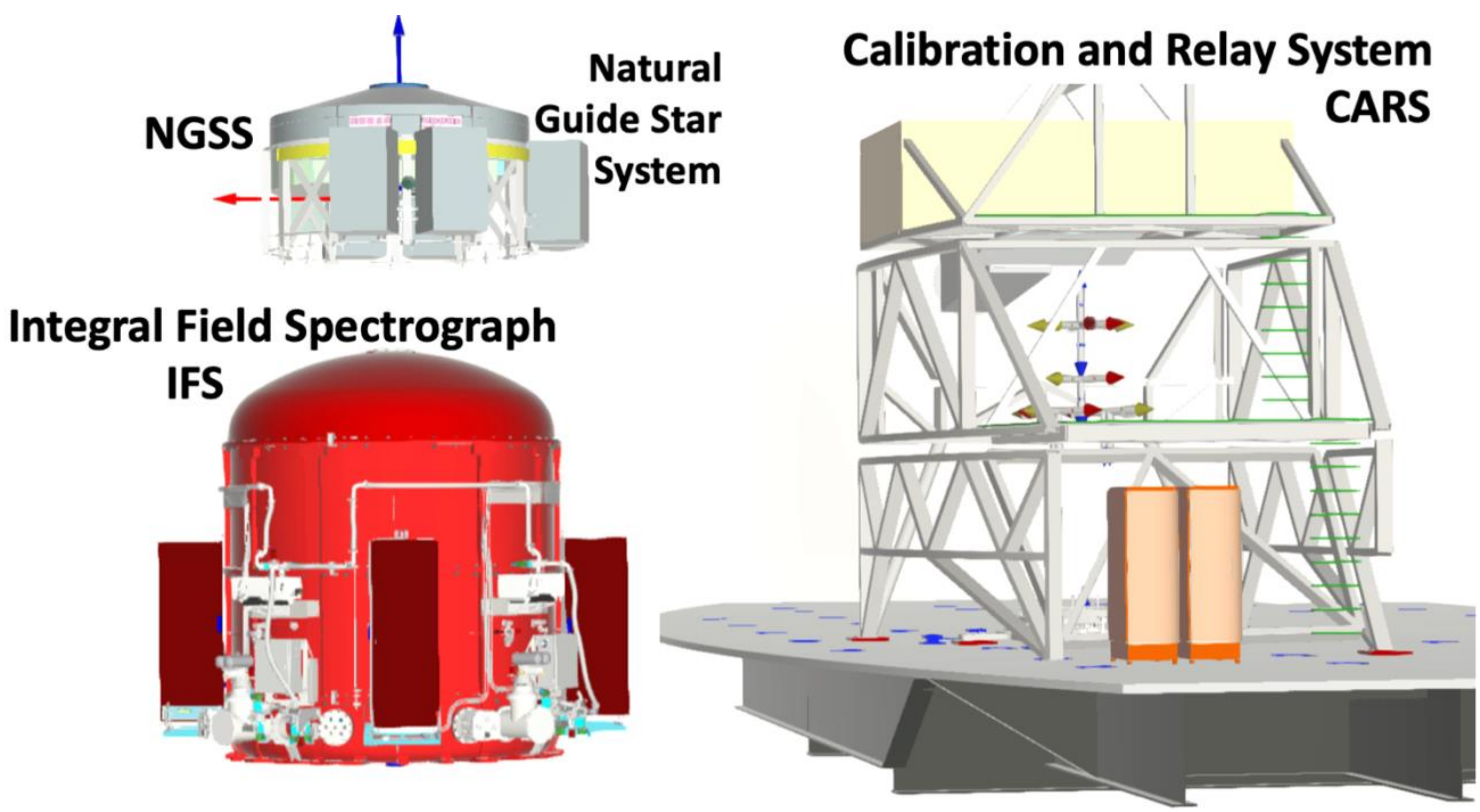

*Figure 6. Overview of the rescoped instrument configurations and fields of view for the two spaxel scales, and the simplified four-channel spectrograph cryostat; the rescope reduces the number of achievable instrument configurations from 181 to 26 and the number of cryogenic mechanisms from 26 to 14.*

**3.1 Calibration and Relay System (CARS)**

The Calibration and Relay System (CARS) is mounted directly beneath MORFEO on the Nasmyth platform and provides the optical interface between the telescope focal plane and the HARMONI cold assembly. Its relay optics, including the M12 mirror, re-image the field delivered by MORFEO onto the HARMONI pre-optics with the pupil and focal-plane conjugations required by the downstream IFS. CARS also embeds a dedicated science calibration unit, used to illuminate the IFS with flat-field and wavelength reference sources for spectral calibration, while the wavefront-sensing calibration sources required to verify the AO correction are provided by MORFEO's own calibration unit.

**3.2 Natural Guide Star Sensing System (NGSS)**

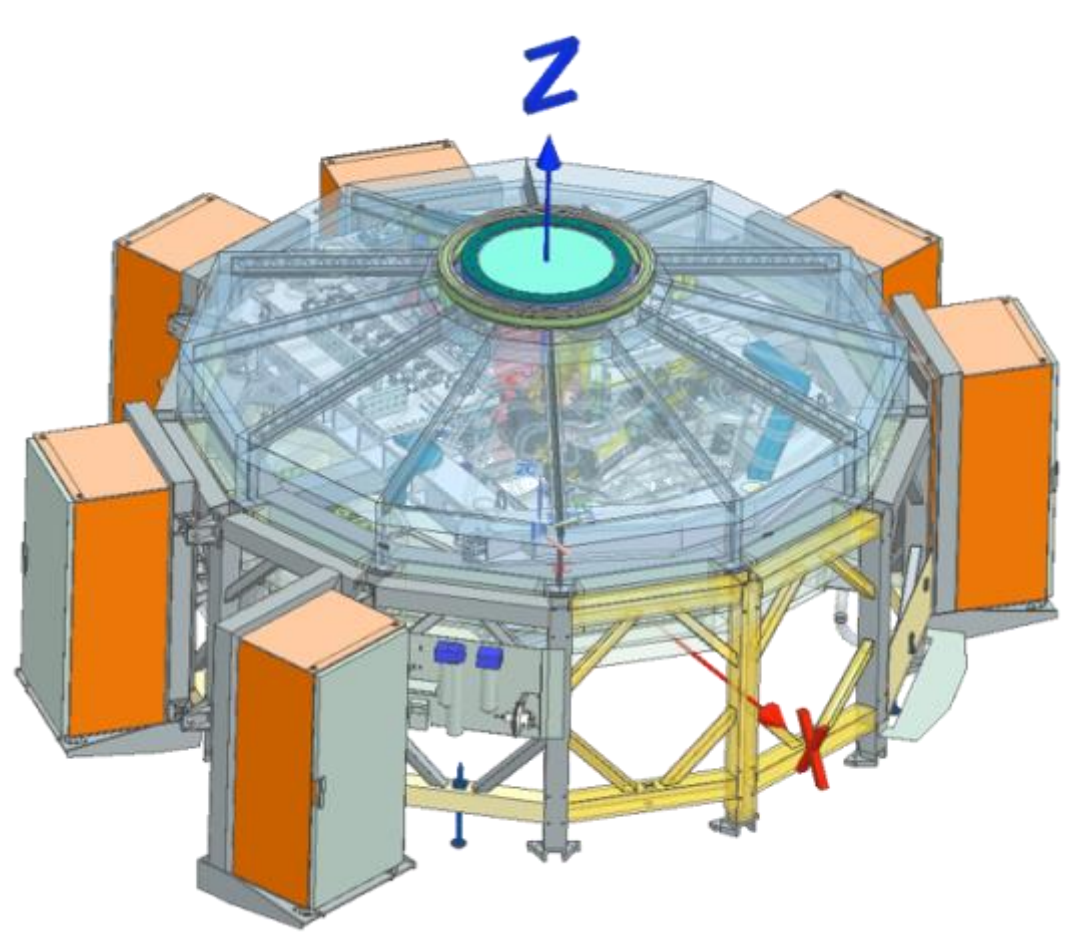

The NGSS is the HARMONI-specific adaptive optics sensing system operating in conjunction with MORFEO in MCAO mode, and independently in SCAO/HCAO mode. It consists of up to three Natural Guide Star (NGS) pick-off arms, each equipped with a low-order wavefront sensing channel, that patrol an annular field of outer diameter 160 arcsec around the IFU centre.[6,7] This patrol field is matched to the MORFEO free field, ensuring coherent sensing across the full MCAO correction volume. In MCAO mode, MORFEO uses the laser guide stars to correct high-order wavefront modes, while HARMONI's NGS arms provide the low-order (tip-tilt and focus) corrections that the laser guide stars cannot sense due to the focal anisoplanatic effect. This division of sensing responsibility between MORFEO and the NGSS is fundamental to achieving the wide sky coverage that is a defining feature of the new HARMONI design.[8]

*Figure 7. View of the Natural Guide Star Sensing System.*

### 3.3 Integral Field Spectrograph (IFS)

The IFS is the scientific heart of HARMONI, converting the two-dimensional field of view delivered by the adaptive optics system into a reformatted pseudo-slit that can be dispersed by the spectrograph. The spatial reformatting is performed by an image slicer, which cuts the field into a series of thin slices and rearranges them end-to-end along a common direction. This approach, long established in ground-based integral field spectroscopy, delivers high optical efficiency and minimal cross-talk between spatial elements.

The IFS delivers two spatial scales. In the fine-scale mode (6×6 mas spaxels), the field of view is 1.23×0.91 arcsec, providing Nyquist sampling of the MORFEO/MCAO PSF in H band across a wide range of natural guide star configurations. In the coarse-scale mode (25×25 mas spaxels), the field of view expands to 5.1x3.8 arcsec, giving access to extended sources and enabling in-IFU sky nodding for many science targets at high redshift. In the High Contrast AO mode, a dedicated scale changer within the coronagraphic module provides a 3.9×3.9 mas spaxel scale over a 0.6×0.8 arcsec field, ensuring Nyquist sampling of the ELT diffraction limit at 1.45 μm.[15,16].

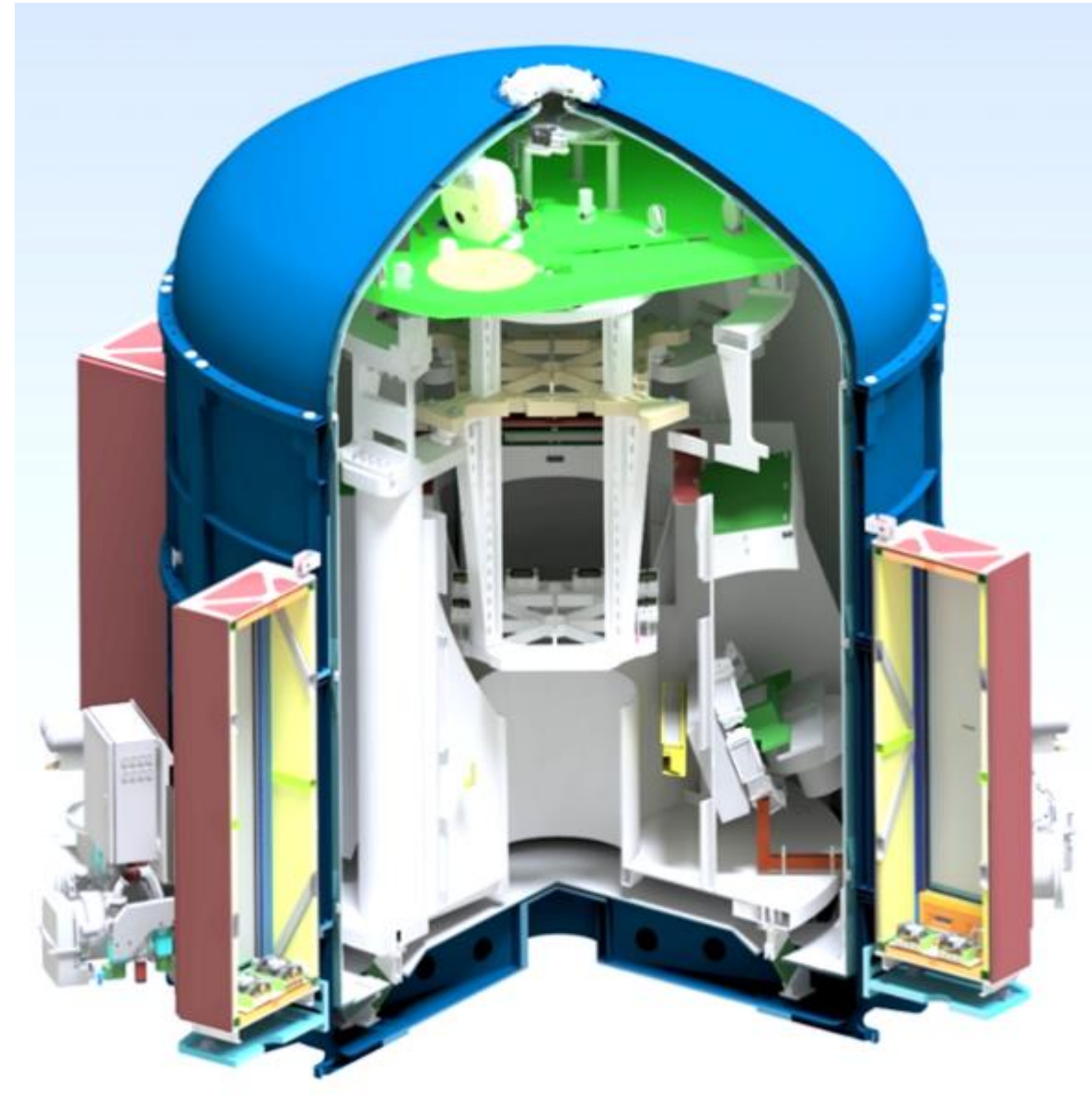

*Figure 8. Cross-section of the HARMONI cryostat and spectrograph assembly, housing the Image Slicer Module, the Splitting and Relay Module, and the four spectrograph channels; overall height 3.5 m.[10,11,12,13,14]*

Dispersed light from each pair of adjacent slices is directed onto a dedicated detector array by an optimised camera lens group. An important design refinement introduced during the rescoping process is the adoption of a grating-dependent anamorphic magnification of up to 15%, applied to individual slices before dispersion. This adjustment addresses the specific challenge of operating at the diffraction limit of a segmented aperture, where diffraction at the pupil edges suppresses the wings of the line spread function (LSF) and can reduce the effective spectral sampling below the Nyquist criterion. The anamorphic correction achieves an average LSF sampling of 2.0 ± 0.05 pixels per FWHM across all wavelengths and both spaxel scales, without adding any mechanical complexity to the cold assembly.[17]

HARMONI provides six spectral settings in two families, as summarised in Table 1. The Medium Resolution family (MR, R≈ 7000) provides four settings covering the Iz, J, H, and K near-infrared windows individually, enabling high-sensitivity spectroscopy between sky emission lines with sufficient velocity resolution to separate galactic kinematic components, detect molecular absorption features, and measure stellar radial velocities at the few-km/s level. The Low Resolution family (LR, R≈ 2750–3000) provides two broad-band settings: Iz+J (0.75–1.35 μm) and H+K (1.45–2.40 μm), optimised for maximum sensitivity across the widest possible wavelength range in a single setting.[18,19]

**Table 1. Spectral configurations of HARMONI.**

| Grating | λref (μm) | R at λref | λmin (μm) | λmax (μm) |
|---|---|---|---|---|
| LR1 (Iz+J) | 1.05 | 3000 | 0.75 | 1.35 |
| LR2 (H+K) | 1.93 | 2750 | 1.45 | 2.40 |
| MR1 (Iz) | 0.94 | 7000 | 0.83 | 1.05 |
| MR2 (J) | 1.19 | 7000 | 1.05 | 1.32 |
| MR3 (H) | 1.63 | 7000 | 1.45 | 1.80 |
| MR4 (K) | 2.19 | 7000 | 1.97 | 2.40 |

At the request of the science working groups, the blue wavelength limit of the LR1 grating has been set to 0.75 µm, the practical short-wavelength limit of near-infrared detector sensitivity and atmospheric transmission, extending coverage beyond the 0.83 µm of earlier designs. This extension, at the cost of a modest reduction in spectral resolution to R~3000 at the blue end, is specifically motivated by kilonova spectroscopy, where the full P-Cygni profile of r-process elements falls in this wavelength range, and by observations of the Calcium triplet in stars and nearby galaxies.

One important addition resulting from the instrument rescope was the addition of a Pointing Camera. The Pointing Camera is a wide-field imaging system mounted adjacent to the IFU, designed to image an area of sky approximately 20 arcsec from the IFU centre simultaneously with science exposures. The PCam serves two critical functions. First, by detecting stars and compact galaxies in the surrounding field during each exposure and cross-matching them against an astrometric reference catalogue, it provides absolute World Coordinate System (WCS) calibration and satisfies the instrument's blind-pointing requirement, to an accuracy better than 30 mas. Second, this absolute registration, combined with the sub-spaxel relative cube-to-cube registration provided by the NGS arms (Section 5.3), enables the combination of data cubes taken on different nights, potentially separated by months, into a coherent final data product, fully exploiting the long-baseline co-addition that is essential for detecting the faintest science targets, and allows cross-matching of HARMONI data cubes with external datasets from JWST, ALMA, and other facilities. This improvement was decisive for the instrument's ability to deliver science at the ELT diffraction limit in long-integration programmes, and it complements the sub-spaxel relative registration required for reliable cube combination.

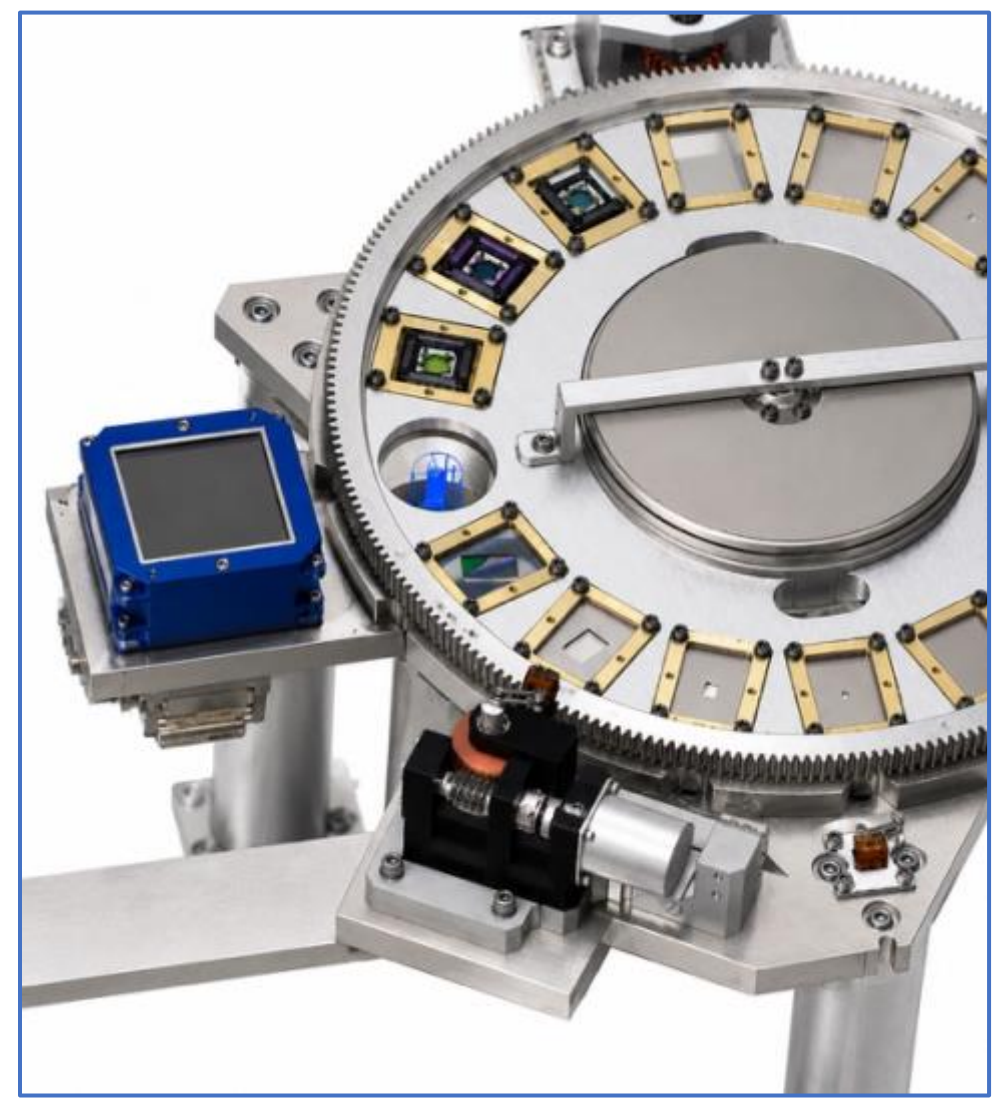

*Figure 9. The Pointing Camera (PCam) module, mounted on the pre-optics scale-changer turret alongside the H4RG focal-plane detector wheel, providing absolute World Coordinate System calibration for HARMONI science exposures.*

## 4. ADAPTIVE OPTICS MODES AND PERFORMANCE

HARMONI operates in three AO modes that together provide a comprehensive range of sky coverage, angular resolution, and contrast performance tailored to different science programs.

### 4.1 Single Conjugate AO (SCAO)

In SCAO mode, HARMONI closes a single-conjugate high-order AO loop using a Natural Guide Star located on or close to the science target, without relying on laser guide stars. This mode delivers the highest possible on-axis Strehl ratio for a given atmospheric condition and guide star brightness, making it the preferred choice for science programmes that require maximum contrast or point-source sensitivity in the immediate vicinity of a bright star. The SCAO loop uses a pyramid wavefront sensor, working down to R=12 and delivering a Strehl ratio of approximately 70% in K band under median atmospheric conditions. [20,21,22]

### 4.2 High Contrast AO (HCAO)

The High Contrast AO mode is designed for the direct detection and spectral characterization of exoplanets and circumstellar disks at small angular separations from a bright star. In HCAO mode, HARMONI uses an on-axis natural guide star to close the SCAO loop and simultaneously passes the science beam through a series of coronagraphic optical elements, an apodizing pupil plane mask and a focal plane coronagraphic mask, that suppress stellar diffraction light both on-axis and at small off-axis angles. A dedicated scale changer within the High Contrast Module changes the plate scale to 3.9×3.9 mas per spaxel, providing Nyquist sampling of the ELT diffraction limit at 1.45 µm.

The primary contrast requirement for HCAO is to detect a point source with a peak intensity ratio of $5\times10^{-5}$ relative to the host star at a separation <100 mas at 1.65 µm, in a single 60-second exposure for guide stars brighter than $m_v = 9$, under the best quartile of atmospheric conditions. This contrast level corresponds to the detection of Jupiter-mass planets at separations of 1–2 AU in systems within 20 pc. As JWST and previous ground-based instruments have demonstrated the power of molecular mapping by cross-correlating observed spectra with atmospheric models, HARMONI/HCAO will extend this technique to planets at separations and masses that are currently inaccessible, enabling composition, temperature, and dynamical constraints on a qualitatively new population of directly imaged planets.[23]

### 4.3 Multi-Conjugate AO with MORFEO (MCAO)

MORFEO is the dedicated MCAO module of the ELT, developed by an INAF-led consortium [24]. It uses six laser guide stars distributed in a constellation around the science field, combined with up to three natural guide stars sensed by HARMONI's NGSS, to correct atmospheric turbulence over the full volume above the telescope pupil. Multi-conjugate correction delivers a more uniform and stable PSF across the HARMONI field of view than single-conjugate correction, and provides substantially wider sky coverage because the low-order correction sensed by the natural guide stars does not require a bright star close to the science target.

For HARMONI+MORFEO, the NGS selection criteria for a typical median-performance observation require one star with H < 17.5 and one or two additional stars with H < 21.0 and R < 22.0 (Vega). Under these conditions, the sky coverage analysis at the South Galactic Pole, the most conservative estimate for extra-galactic science programmes, demonstrates that more than 75% of the sky is accessible at Strehl ratios above 50% in K band under good conditions. The sky coverage rises to above 90% of accessible fields at 30% Strehl.[25,26]

The MCAO performance can be explored using TipTop.[27]

## 5. EXPECTED PERFORMANCE

### 5.1 Point source sensitivity

The point-source continuum sensitivity of HARMONI has been evaluated as a function of wavelength for the MR and LR grating families, adopting the HARMONI+MORFEO configuration, 10 ks exposures, and a detection threshold of 10σ per spectral pixel. Under median conditions (0.67 arcsec natural seeing, airmass 1.1) and the 6 mas spaxel scale, HARMONI reaches AB~23.5–24.0 per spectral pixel across the I, J, H, and K bands in the R~7000 configuration, and AB~25 in the H+K LR configuration at its most sensitive wavelengths between sky lines. For direct comparison with JWST/NIRSpec at the same spectral resolution (R~2700), HARMONI is 1.5–2 magnitudes more sensitive for point sources, equivalent to a factor 15–60 reduction in the exposure time required to achieve a given signal-to-noise ratio, at 6 times the angular resolution.

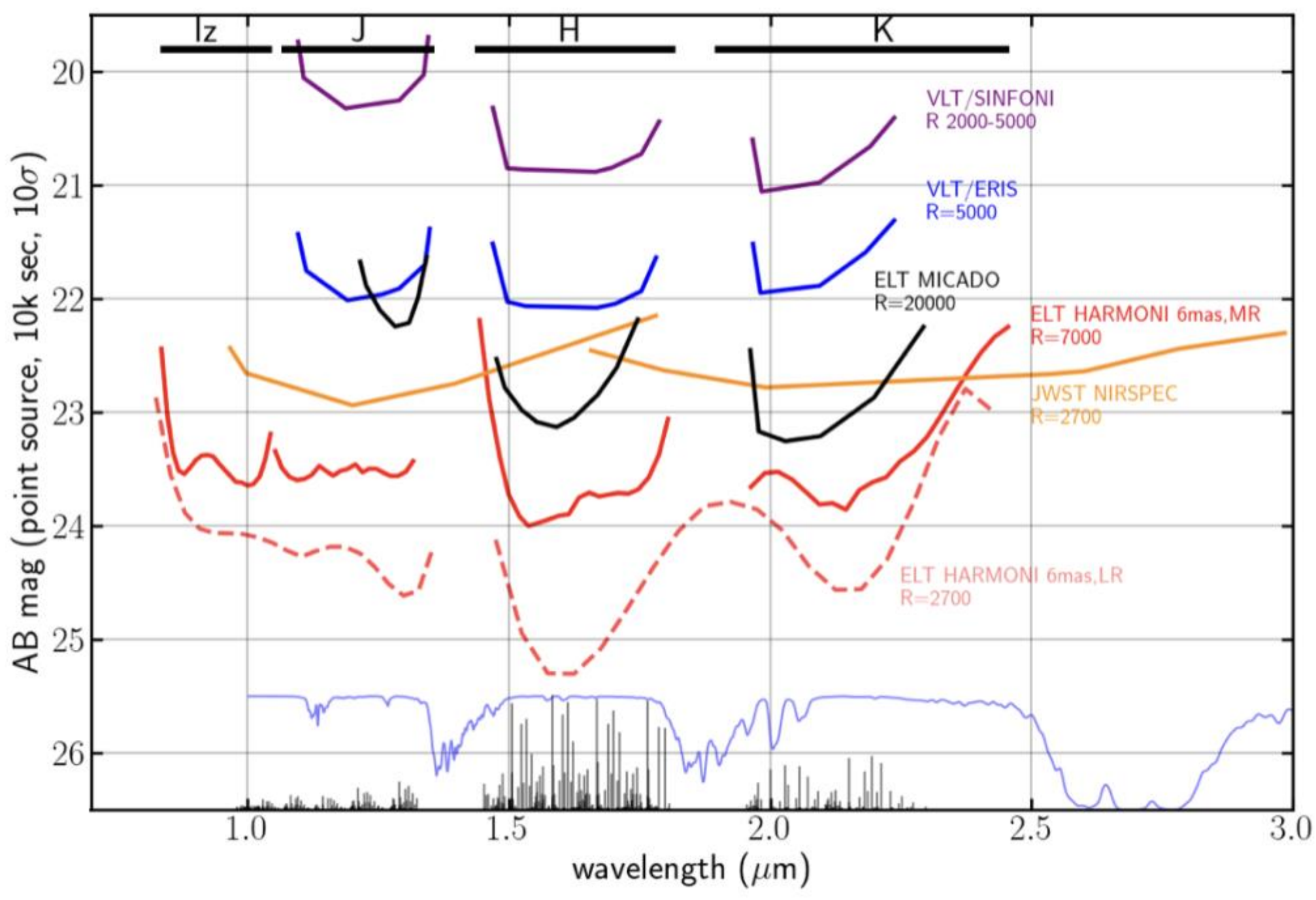

*Figure 10. Point-source continuum sensitivity (10σ detection, 10 ks integration) as a function of wavelength for HARMONI+MORFEO, compared to VLT/SINFONI, VLT/ERIS, ELT/MICADO, and JWST. Thanks to the collecting power of the 39 m ELT, HARMONI is expected to be several tens of times more sensitive than JWST.*

### 5.2 PSF FWHM and image quality in long integrations

For science programmes that require the combination of multiple Observing Blocks (OBs) to accumulate sufficient signal-to-noise, the effective PSF in the final stacked data cube is the key quality metric. In the 6 mas scale, under median atmospheric conditions with a typical 3-NGS MORFEO asterism, the expected PSF FWHM values are summarised in Figure 11.

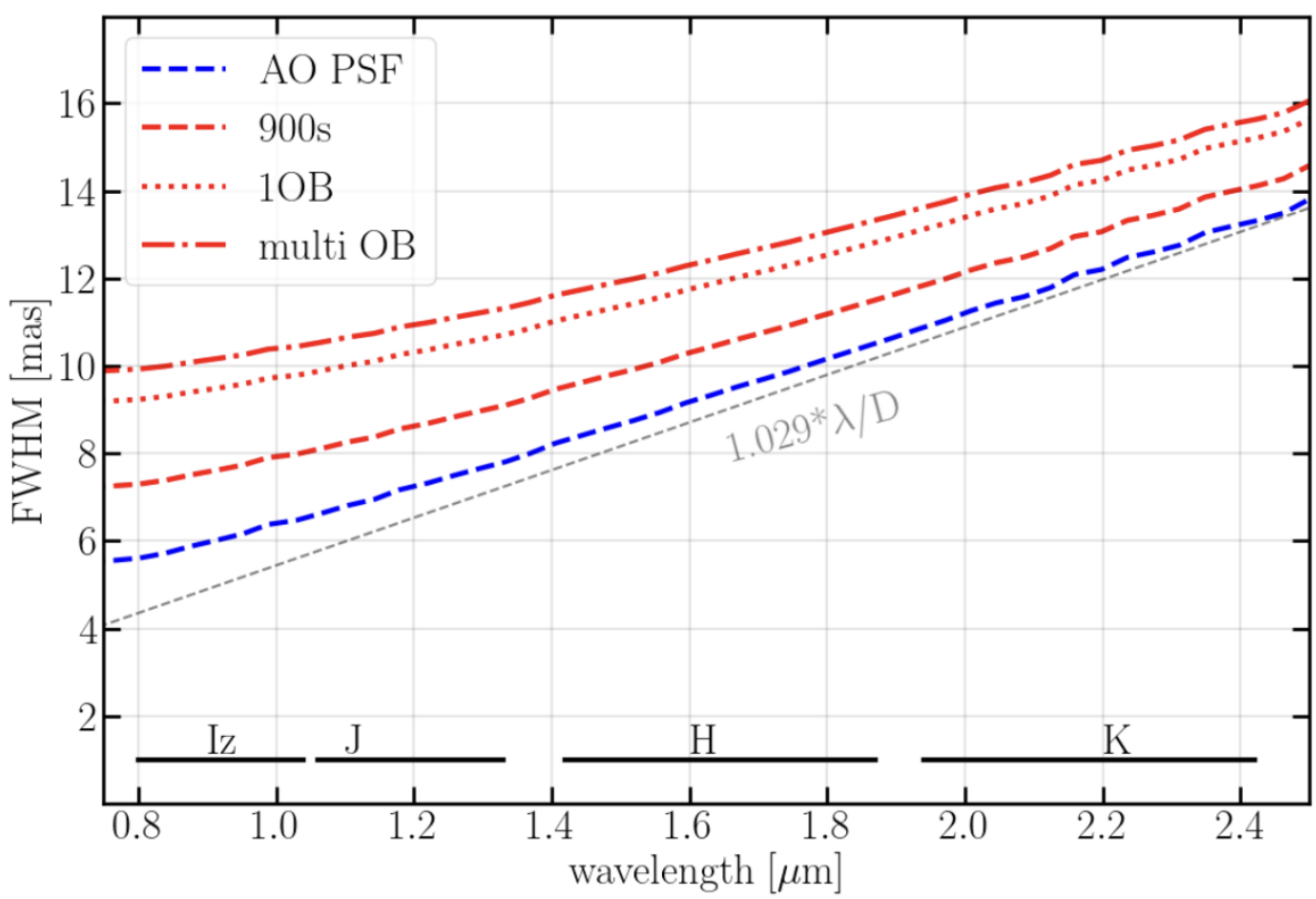

*Figure 11. Predicted PSF FWHM as a function of wavelength for a single AO-corrected exposure, a single 900 s exposure, a single Observing Block, and a multi-OB co-added exposure, compared to the ELT diffraction limit (1.029 λ/D).*

Between the instantaneous AO-corrected PSF and the 900 s exposure PSF, an additional blur is introduced by residual field distortion in the NGS focal plane downstream of the ELT and MORFEO; this effect is referred to as PSF drift. A further jitter term is added when combining independent exposures within a full Observing Block, since the exact field position varies slightly from one exposure to the next. Finally, the quality of the multi-OB co-added PSF is primarily set by the accuracy of the Pointing Camera in re-aligning images acquired over observing baselines of several months. This level of PSF stability and WCS repeatability is essential for HARMONI's most ambitious science programmes, including the detection of faint sources and the mapping of galaxy kinematics at the highest spatial resolution.

### 5.3 Astrometric performance and WCS accuracy

The ability to reliably register data cubes taken in different observing blocks, potentially separated by weeks or months, and under different seeing and thermal conditions, is a fundamental operational requirement for the most demanding HARMONI science cases. HARMONI achieves this through the combination of the NGS arms and the Pointing Camera. The three NGS arms provide relative cube-to-cube registration with an rms accuracy of better than 0.5 PSF FWHM (approximately 3–5 mas in H band for a typical multi-OB programme), enabling blind co-addition to better than 1 spaxel and thereby avoiding significant degradation of the co-added PSF. The PCam provides absolute astrometric calibration with an accuracy better than 30 mas, enabling the cross-matching of HARMONI data cubes with external datasets from JWST, ALMA, and other facilities.

## 6. PROJECT ORGANISATION AND SCHEDULE

### 6.1 Leadership and Project Management Office

The HARMONI project is led by a new Principal Investigator, Professor Jim Dunlop from the University of Edinburgh, who formally took up the role on 1 July 2025. The Project Management Office (PMO) also comprises Dr. Benoit Neichel (LAM) as deputy-PI, Professor Mark Swinbank (Durham University) as Project Scientist, Dr. Adriana de Lorenzo-Cáceres Rodríguez (IAC) as Deputy Project Scientist, Dr. William Taylor (UK Astronomy Technology Centre) as Instrument Scientist, Pr. Matthias Tecza as deputy Instrument Scientist (University of Oxford), Bethany Johnson as System Engineer, and Anne Costille (LAM) as deputy System Engineer. The Consortium Project Manager is Stephen Chittick (UK-ATC), working with David Le Mignant (LAM) as Deputy-PM. They both led the rescope process since its inception in October 2024.

The project is governed by an Executive Board composed of Directors of consortium institutes and the ESO Instrumentation Programme Manager. Consortium Co-Investigators are embedded in ten institutes across Europe and the United States, each carrying responsibility for defined work packages within the instrument. The Science Working Groups, covering stellar populations, nearby galaxies and black holes, the high-redshift universe, the Solar System, exoplanets, and time-domain astronomy, provide the scientific requirements and prioritisation that drive the technical design.

### 6.2 Consortium institutes and work package responsibilities

The HARMONI consortium brings together twelve institutes with complementary expertise in instrumentation, adaptive optics, software, and science. The lead institutes are:

- **UK Astronomy Technology Centre (UK-ATC, STFC)**
- **Laboratoire d'Astrophysique de Marseille (LAM, CNRS/AMU)**
- **Durham University**
- **University of Oxford**
- **University of Edinburgh**
- **Instituto de Astrofísica de Canarias (IAC)**
- **Centro de Astrobiología (CAB)**
- **Centre de Recherche Astrophysique de Lyon (CRAL)**
- **Institut de Planétologie et d'Astrophysique de Grenoble (IPAG)**
- **University of Michigan**
- **Institut de recherche en Astrophysique et Planétologie (IRAP)**
- **ONERA (The French Aerospace Lab)**

### 6.3 Schedule and path to delivery

Following the ESO Science Technical Committee (STC) meeting of April 2026, the rescoped HARMONI concept and its new technical specifications were endorsed by the STC and subsequently approved by the ESO Council at its meeting of 9-10 June 2026, formally clearing the project to resume business as usual. A bottom-up project schedule has been constructed with the IFS Assembly, Integration and Test (AIT) activity on the critical path. A Rescope Design Review (RDR) is scheduled for November 2026 to formally ratify the new technical specifications and confirm compliance of all subsystems before subsystem-level Critical Design Reviews begin. Some subsystems, such as the Image Slicer Module, have undergone minimal design changes during the rescope and procurement of long-lead optics has already commenced, while others, notably the spectrograph, have been substantially redesigned and require full optical CDR cycles.

The current baseline schedule targets a Preliminary Acceptance Europe (PAE) milestone in November 2035. A set of pull-forward options is under active investigation: parallel assembly of the four spectrograph channels using a second assembly team and cryostat, early cold-testing of the completed IFU prior to integration into the full IFS assembly (potentially removing one thermal cycle from the AIT sequence), and the procurement of a first spectrograph unit for prototype testing on an accelerated schedule. The aim is to pull the PAE date to 2034.

From a resources perspective, the rescoped instrument requires a total hardware cost of approximately €31.2M from the consortium (excluding ESO-supplied detectors). The total FTE effort is estimated at approximately 1,400 person-years across the full project lifetime.

## 7. CONCLUSIONS

HARMONI is positioned to be one of the defining scientific instruments of the coming decade. As the integral field spectrograph for the ELT, it will deliver spatially resolved near-infrared spectroscopy at a level of angular resolution, sensitivity, and astrometric stability that has no parallel among existing or planned facilities. From the characterisation of exoplanet atmospheres at sub-AU separations to the mapping of galaxy kinematics at the epoch of reionisation, HARMONI addresses questions that go to the heart of contemporary astrophysics.

The instrument design has been thoroughly reviewed and streamlined through the rescope process: it is technically compliant, within budget, and deliverable on a realistic schedule. The new architecture preserves and in key respects enhances the two defining scientific capabilities: diffraction-limited imaging with wide sky coverage, and full near-infrared spectral coverage at intermediate resolution. The rescoped design also significantly reduced sources of technical risk that would have jeopardised delivery. With a renewed leadership team, a focused and experienced consortium, and a clear path to first light in the second half of the 2030s, HARMONI is ready to take the next step toward transforming our view of the Universe.

## REFERENCES

[1] A. MacIver, “HARMONI at ELT: a dynamic systems engineering approach for rescoping the HARMONI integral field spectrograph,” Proc. SPIE 14152-32, Modeling, Systems Engineering, and Project Management for Astronomy XII (2026).

[2] A. Calderhead, A. MacIver, et al., “HARMONI at ELT: developing a modular review process using TRIZ,” Proc. SPIE 14152-107, Modeling, Systems Engineering, and Project Management for Astronomy XII (2026).

[3] J. Kormendy and L. C. Ho, “Coevolution (Or Not) of Supermassive Black Holes and Host Galaxies,” Annu. Rev. Astron. Astrophys. 51, 511–653 (2013).

[4] N. M. Förster Schreiber and S. Wuyts, “Star-Forming Galaxies at Cosmic Noon,” Annu. Rev. Astron. Astrophys. 58, 661–725 (2020).

[5] R. P. Naidu, et al., “A Cosmic Miracle: A Remarkably Luminous Galaxy at $z_{spec}$=14.44 Confirmed with JWST,” The Open Journal of Astrophysics 9, id. 56033 (2026).

[6] G.J.C. Carballal, I.M.F. Rodríguez, et al., “HARMONI at ELT: LPOU calibrations for a micron-level pointing accuracy,” Proc. SPIE 14154-367, Advances in Optical and Mechanical Technologies for Telescopes and Instrumentation VII (2026).

[7] G.M. Rubio, I.F. Vecino, et al., “HARMONI at ELT: design & tests on the current status of low order wavefront subsystem pick off arm engineering model (LPOA-EM),” Proc. SPIE 14154-368, Advances in Optical and Mechanical Technologies for Telescopes and Instrumentation VII (2026).

[8] K. Dohlen, D. Le Mignant, and A. Bonnefoi et al., “HARMONI at ELT: reoptimization of the natural guide star sensors (NGSS) system for MCAO compatibility,” Proc. SPIE 14150-371, Adaptive Optics Systems X (2026).

[9] J. Migniau, A. Jeanneau, et al., “HARMONI at ELT: cryogenic mirror mount based on bonded Invar supports,” Proc. SPIE 14154-363, Advances in Optical and Mechanical Technologies for Telescopes and Instrumentation VII (2026).

[10] A. Fitzpatrick, J. Mann, et al., “HARMONI at ELT: cryogenic lens mount prototyping – conclusions for the ICAM subsystem,” Proc. SPIE 14154-285, Advances in Optical and Mechanical Technologies for Telescopes and Instrumentation VII (2026).

[11] V. Mauger-Vauglin, and J. Migniau et al., “HARMONI at ELT: update and numerical simulation of a cryostat for HARMONI AIT,” Proc. SPIE 14154-401, Advances in Optical and Mechanical Technologies for Telescopes and Instrumentation VII (2026).

[12] M.E. Cisneros-González, J. Kariuki, et al., "HARMONI at ELT: repeatability analysis of the spectrograph dispersers wheel mechanism," Proc. SPIE 14154-370, Advances in Optical and Mechanical Technologies for Telescopes and Instrumentation VII (2026).

[13] K. McCall, E. Castillo-Domínguez, et al., "HARMONI at ELT: evaluation of opto-mechanical bonding: adhesive performance, substrate compatibility, and transportation viability," Proc. SPIE 14154-369, Advances in Optical and Mechanical Technologies for Telescopes and Instrumentation VII (2026).

[14] R.M. Hernandez, Y. Martin, et al., "HARMONI at ELT: laboratory tests for optimization and validation of the scale changer turret mechanism," Proc. SPIE 14149-288, Ground-based and Airborne Instrumentation for Astronomy XI (2026).

[15] A.A.L. González, E.H. Suárez, et al., "HARMONI at the ELT: sub-micron repeatability in the cryogenic scale-changer turret for pre-optics: design, optimization, and test results," Proc. SPIE 14154-99, Advances in Optical and Mechanical Technologies for Telescopes and Instrumentation VII (2026).

[16] S.P. Todd, É. Harvey, et al., "HARMONI at ELT: line spread functions in a diffraction limited spectrometer," Proc. SPIE 14154-71, Advances in Optical and Mechanical Technologies for Telescopes and Instrumentation VII (2026).

[17] E.R. Muslimov, M. Tecza, et al., "HARMONI at ELT: optical design of the spectrograph sub-system," Proc. SPIE 14149-298, Ground-based and Airborne Instrumentation for Astronomy XI (2026).

[18] P. Bañares-Palacios et al., "HARMONI at ELT: integral field spectrograph pre-optics (IPO) design after rescope process," Proc. SPIE 14149-290, Ground-based and Airborne Instrumentation for Astronomy XI (2026).

[19] J.D. Floriot, T. Pichon, and K. Dohlen et al., "HARMONI at ELT: SCAOS design and development plan after rescope," Proc. SPIE 14149-217, Ground-based and Airborne Instrumentation for Astronomy XI (2026).

[20] J.D. Floriot, et al., "HARMONI at ELT: prototype developments for the SCAOS object-selection and dichroic exchange modules," Proc. SPIE 14154-182, Advances in Optical and Mechanical Technologies for Telescopes and Instrumentation VII (2026).

[21] C.Z. Bond et al., "HARMONI at ELT: from simulations to on-sky prototyping of the HARMONI single conjugate AO system," Proc. SPIE 14150-77, Adaptive Optics Systems X (2026).

[22] A. Carlotti, L. Jocou, et al., "HARMONI at ELT: from design trade-offs to prototypes of the high-contrast capability," Proc. SPIE 14149-230, Ground-based and Airborne Instrumentation for Astronomy XI (2026).

[23] P. Ciliegi, et al., "MAORY/MORFEO at ELT: general overview up to the preliminary design and a look towards the final design," Proc. SPIE 12185, 1218514 (2022).

[24] J. Sauvage et al., "HARMONI at ELT: status of the HARMONI MCAO mode," Proc. SPIE 14150-74, Adaptive Optics Systems X (2026).

[25] L.G. Staykov, T.J. Morris, and D.C. Malone, "HARMONI at ELT: pupil motion tracking and compensation using a wave-front sensor," Proc. SPIE 14150-112, Adaptive Optics Systems X (2026).

[26] B. Neichel, O. Beltramo-Martin, C. Plantet, et al., "TIPTOP: a new tool to efficiently predict your favorite AO PSF," Proc. SPIE 11448, Adaptive Optics Systems VII, 114482T (2021).

[27] https://www.harmoni-elt.eu/